\documentclass[slac_one]{revtex4}
\usepackage{graphicx}
\usepackage{fancyhdr}
\usepackage{subfigure}
\begin{document}

\title{Prospects for early top anti-top resonance searches in ATLAS} 

%

\author{Bertrand Chapleau, on behalf of the ATLAS collaboration}
\affiliation{McGill University, Montreal, Canada}

\begin{abstract}
Searches for signatures of new physics in top anti-top events at the LHC require efficient reconstruction of top quarks with a broad range of transverse momenta. Three new reconstruction schemes are developed to deal with the large variety of top decay topologies. Their performance on the lepton + jets final state is evaluated using a detailed simulation of signal and background processes. Compared to previous ATLAS studies, a much improved reconstruction efficiency is found over a large top anti-top invariant mass range. As a consequence, even in the earliest phase of the experiment, ATLAS is expected to significantly extend the mass reach of existing searches.
\end{abstract}

\maketitle

\thispagestyle{fancy}





\section{BOOSTED $t \bar t$ TOPOLOGIES}

\begin{figure}[h]
\centering
\subfigure[] { \includegraphics[width=0.3\textwidth]{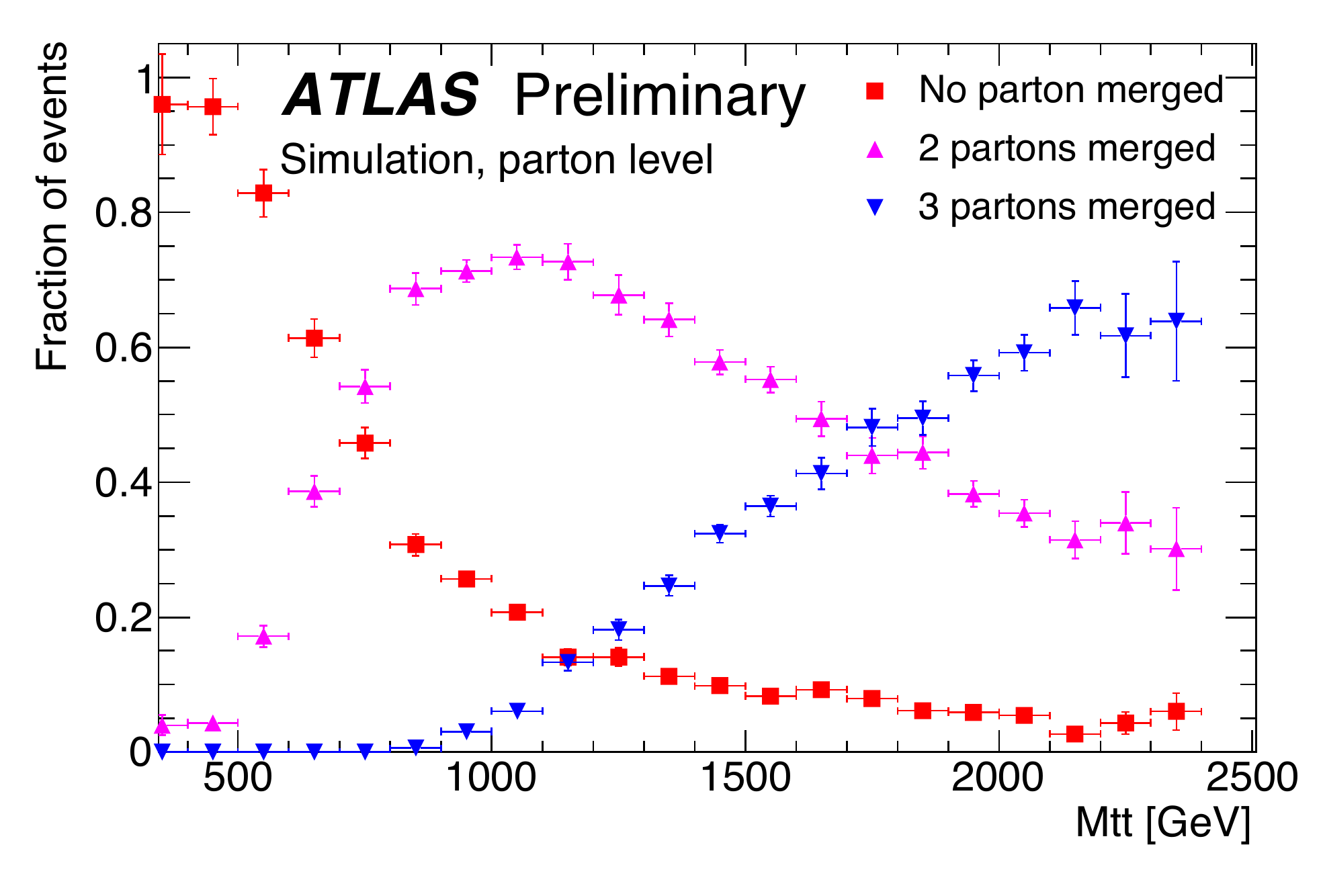} }
\subfigure[] { \includegraphics[width=0.325\textwidth]{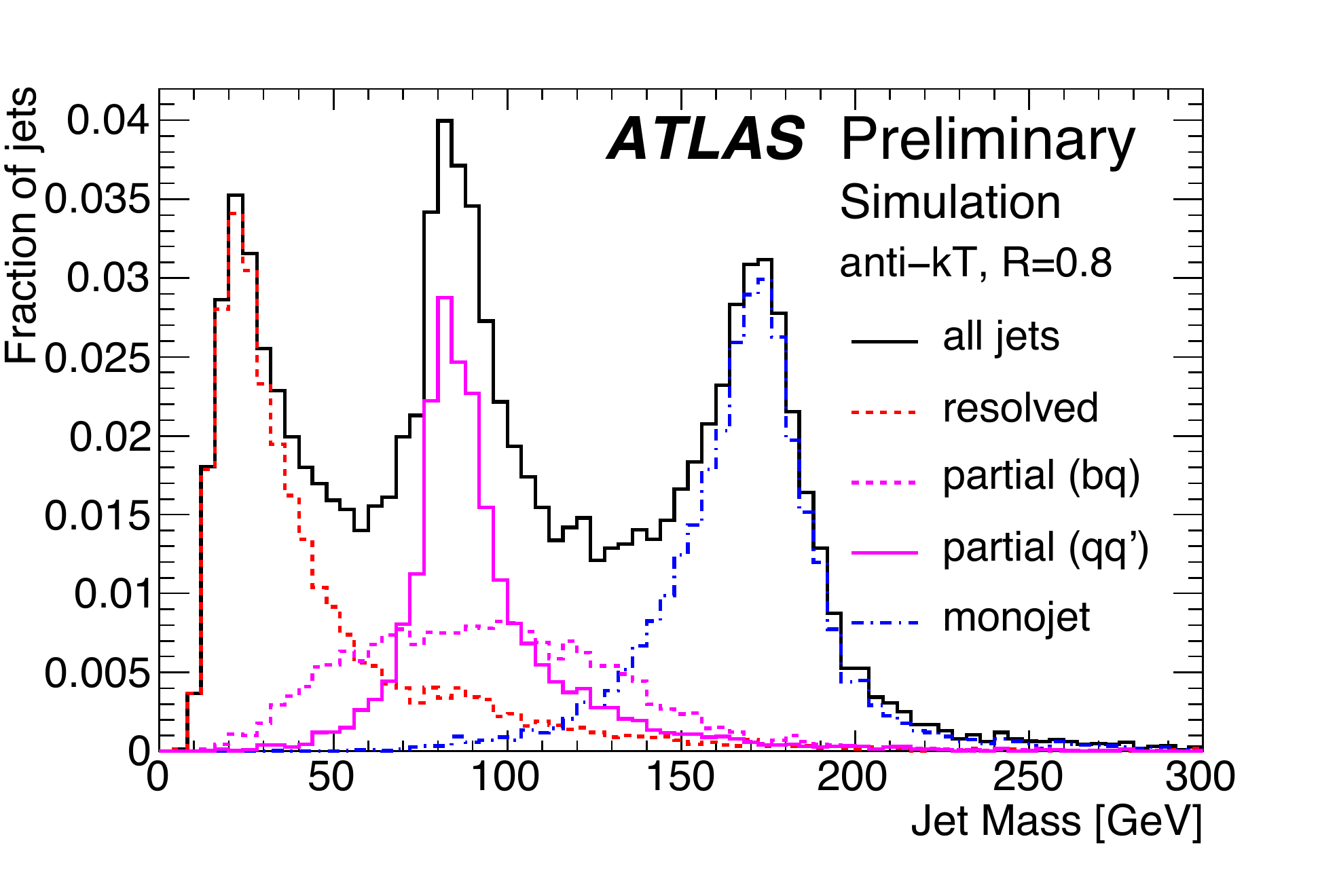} }
\caption{a) Probability that partons from a hadronic top decay are found within a $\Delta R = \sqrt{\Delta \eta^2 + \Delta \phi^2}$ distance of 0.8. b) Reconstructed invariant mass of
the leading jet in $pp \rightarrow X \rightarrow t \bar t \rightarrow$ lepton+jets events. }
\label{topo_jetmass}
\end{figure}

One of the most challenging aspects of heavy $t \bar t$ resonance searches lies in
the reconstruction and identification of boosted top quark decays. A top
quark being produced with very high transverse momentum is a source of a
new experimental phenomenology: its decay products become very
collimated (Figure~\ref{topo_jetmass}.a) and as a consequence, jets tend to merge into a single reconstructed jet.
Different boost regimes will give rise to different event topologies. As shown in Figure~\ref{topo_jetmass}.b the mass
of the heaviest jet (the jet mass is defined as the mass of the 4-momenta sum of its massless constituents) in the event can be used to classify such topologies.

\section{RECONSTRUCTION ALGORITHM}

\begin{figure}[h]
\centering
\subfigure[] { \includegraphics[width=0.317\textwidth]{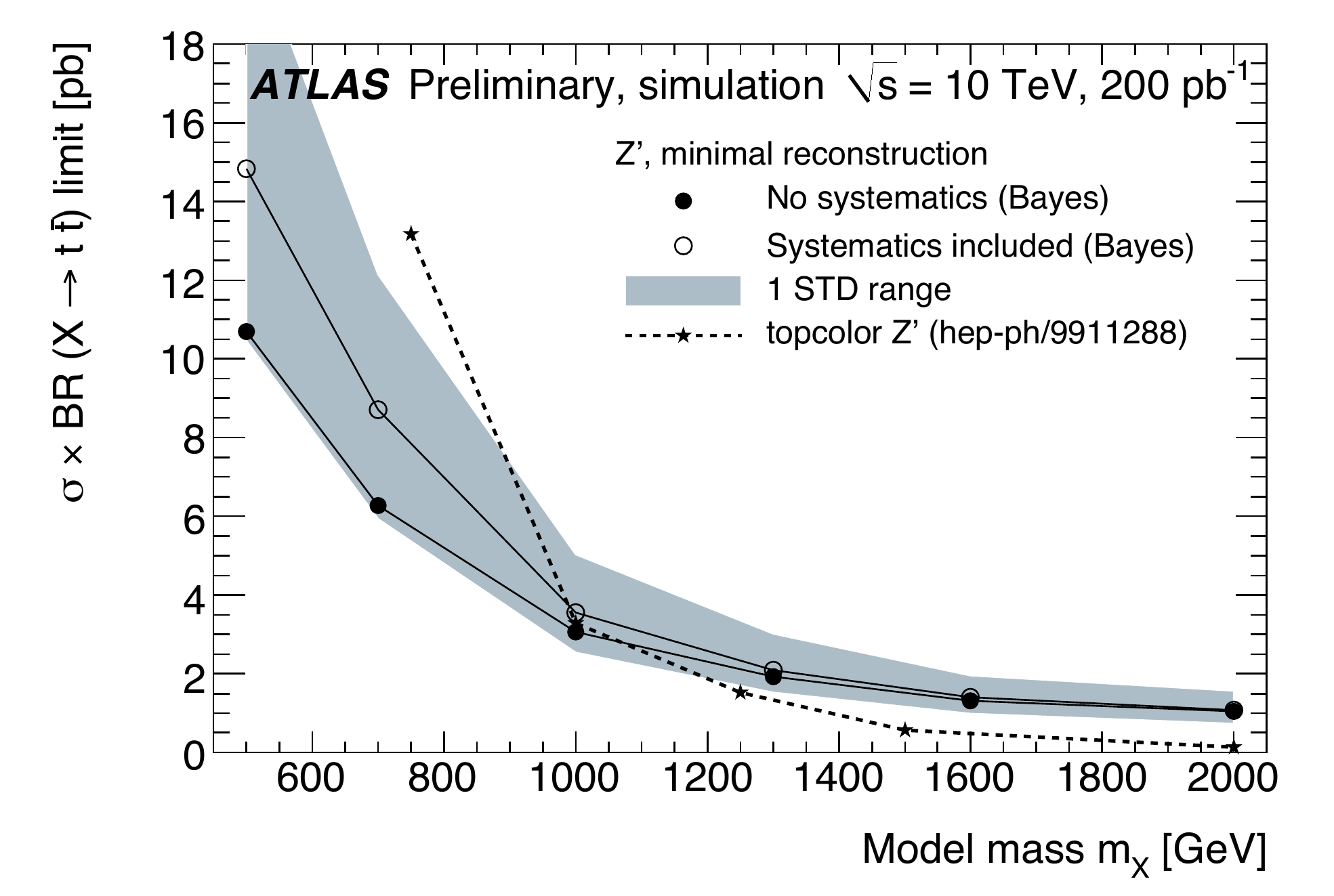} }
\subfigure[] { \includegraphics[width=0.317\textwidth]{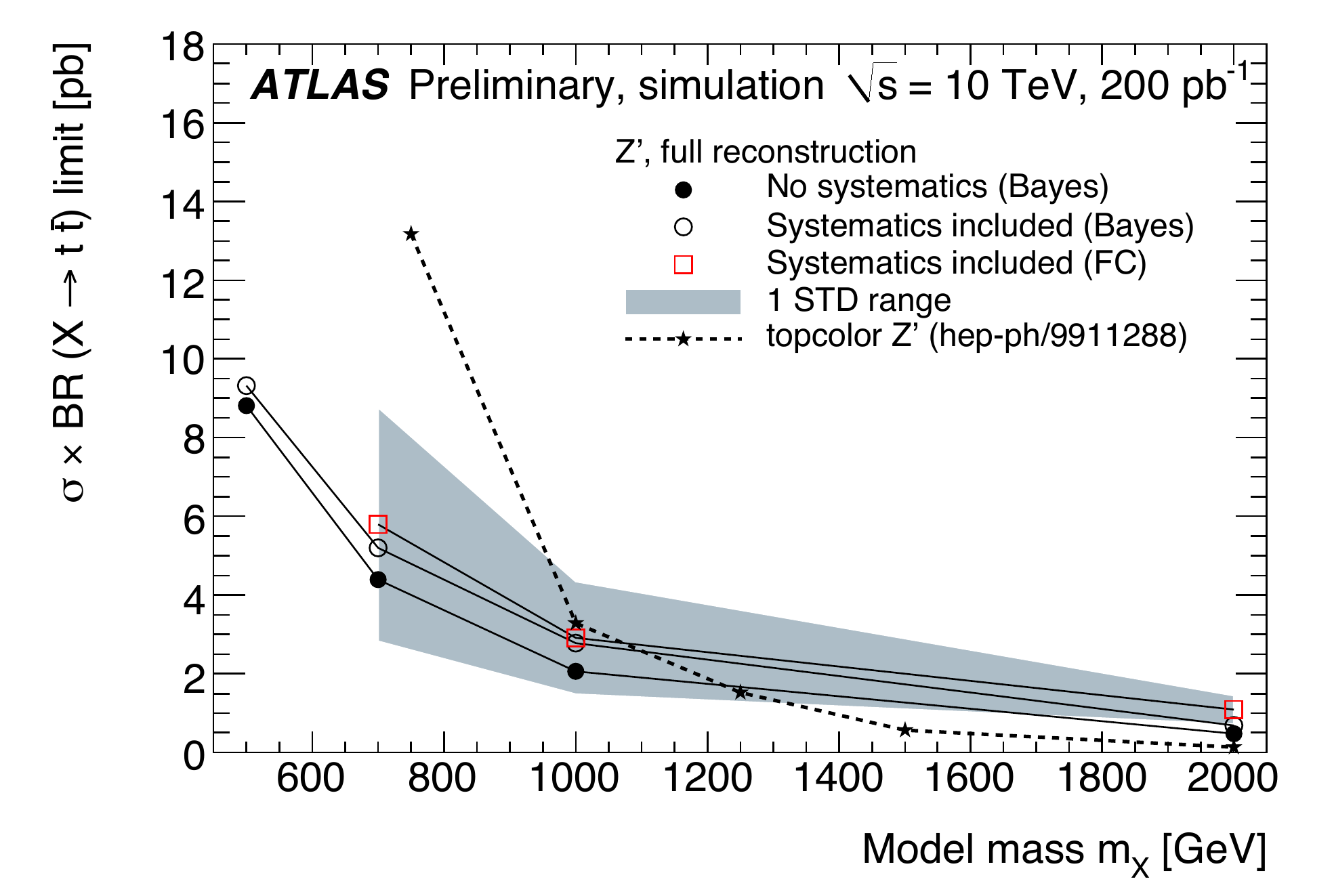} }
\subfigure[] { \includegraphics[width=0.317\textwidth]{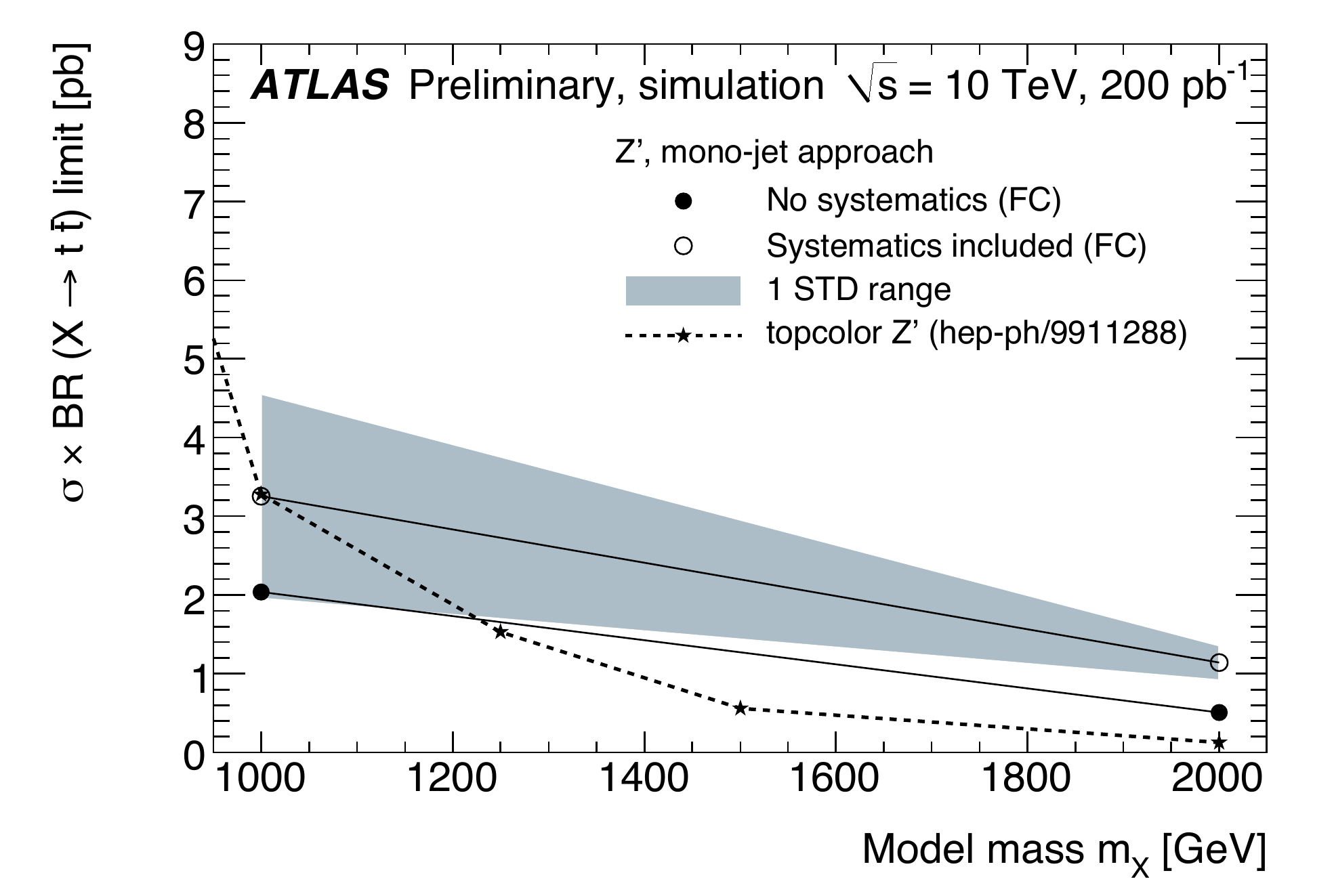} }
\caption{ATLAS sensitivity projection (95 \%
C.L. signal cross-section limit)
for a narrow resonance obtained from the different reconstruction approaches after 200 pb$^{-1}$ at center-of-mass energy of 10 TeV: a) minimal, b) full and c) mono-jet approach. The dashed line corresponds to the production cross section times branching fraction into $t\bar t$ of a Topcolor $Z'$ resonance.}
\label{sensitivities}
\end{figure}

\subsection{Minimal reconstruction}

The main motivation that drives the so-called minimal approach is  its applicability in the very early stages of the experiment with a high signal efficiency over a wide range of $m_{t \bar t}$. It relies on a few number of observables, no flavour tagging (for $b$-jets) is employed and it does not attempt to reconstruct the top quarks individually. As a consequence, the algorithm can be commissioned early on in the experiment and the sensitivity to systematic biases is minimised. 

Jets are defined by means of the ATLAS Cone algorithm with a $R$ parameter of 0.4 and only those jets with transverse energy $E_{T}$ greater than 40 GeV are further considered in the analysis. The events are classified according to the highest jet mass and the number of jets in the event. Thus with 4 or more selected jets, $m_{t \bar t} = m_{jjjjl\nu}$, where $jjjj$, $l$ and $\nu$ refers to the 4 highest $E_{T}$ jets, the reconstructed letpon and the reconstructed neutrino (from the $W$ mass constraint using the missing $E_T$ and the lepton momentum), respectively. With 3 selected jets, similarly, $m_{t \bar t} = m_{jjjl\nu}$ but the sample is further split into two subsamples of events according to the highest jet mass found in the event (that is if $m_j^{max}$ is above or below 65 GeV). In the absence of signal, an exclusion limit of $\sigma \times$ BR($X \rightarrow t \bar t$) = 3.6 pb (1.1 pb) is expected with this method for a resonance mass of 1 TeV (2 TeV) (see Figure~\ref{sensitivities}.a).

\subsection{Full reconstruction}

This algorithm aims at providing a much tighter control of the reducible background by performing, as its name suggests, a full reconstruction of the top and anti-top quark. Flavour tagging is also used. 

Jets are defined by means of the Anti-$k_{\perp}$ algorithm with a $R$ parameter of 0.4 and only those jets with transverse energy $E_{T}$ greater than 20 GeV are further considered in the analysis. This full reconstruction approach also adapts its reconstruction scheme according to the event topology which is determined by classifying events according to the highest jet mass in the event. The measured top masses are in addition replaced by the generated top mass $m_{t}^{PDG}$ in order to improve the resonance mass resolution. In the absence of signal, an exclusion limit of $\sigma \times$ BR($X \rightarrow t \bar t$) = 2.9 pb (1.1 pb) is expected with this method for a resonance mass of 1 TeV (2 TeV) (see Figure~\ref{sensitivities}.b).
\begin{itemize}
\item  $m_j^{max} < 65$ GeV (resolved) : \\  $>= 4$ jets required, among which 2 should be identified $b$-jets. $m_{Z'} = m_{bjjbl\nu} - m_{bjj} - m_{bl\nu} + 2 m_{t}^{PDG}$
\item 65 GeV $< m_j^{max} < 130$ GeV (partial merge) : \\  $>= 3$ jets required, among which 1 should be identified $b$-jets.  $m_{Z'} = m_{jjbl\nu} - m_{jj} - m_{bl\nu} + 2 m_{t}^{PDG}$
\item $m_j^{max} > 130$ GeV (mono-jet) : \\  $>= 2$ jets required, among which 1 should be identified $b$-jets.  $m_{Z'} = m_{jbl\nu} - m_{j} - m_{bl\nu} + 2 m_{t}^{PDG}$
\end{itemize}

\subsection{The mono-jet reconstruction approach}

The mono-jet reconstruction algorithm favors the high end of the $m_{t \bar t}$ spectrum where top quarks are highly boosted. It offers a good mass resolution and a strong handle on the background processes. As opposed to the previous two reconstruction approaches, this algorithm solely relies on the mono-jet topology which is enhanced by the choice of the jet definition. Indeed, the Anti-$k_{\perp}$ algorithm with a $R$ parameter of 1.0 is employed. Three-dimensional topological calorimeter clusters are used as inputs and only those jets with transverse energy $E_{T}$ greater than 200 GeV are further considered in the analysis to reflect the assumption that the top quarks are boosted. Only at least two jets are required and $m_{Z'} = m_{j,jl\nu}$, wjere $j$ is the identified top mono-jet and $jl\nu$ is the reconstructed semi-leptonic top.  In the absence of signal, an exclusion limit of $\sigma \times$ BR($X \rightarrow t \bar t$) = 3.3 pb (1.1 pb) is expected with this method for a resonance mass of 1 TeV (2 TeV) (see Figure~\ref{sensitivities}.c). The boosted top identification is detailed in the following sub-sections. \\

\begin{figure}[h]
\centering
\subfigure[] { \includegraphics[width=0.23\textwidth]{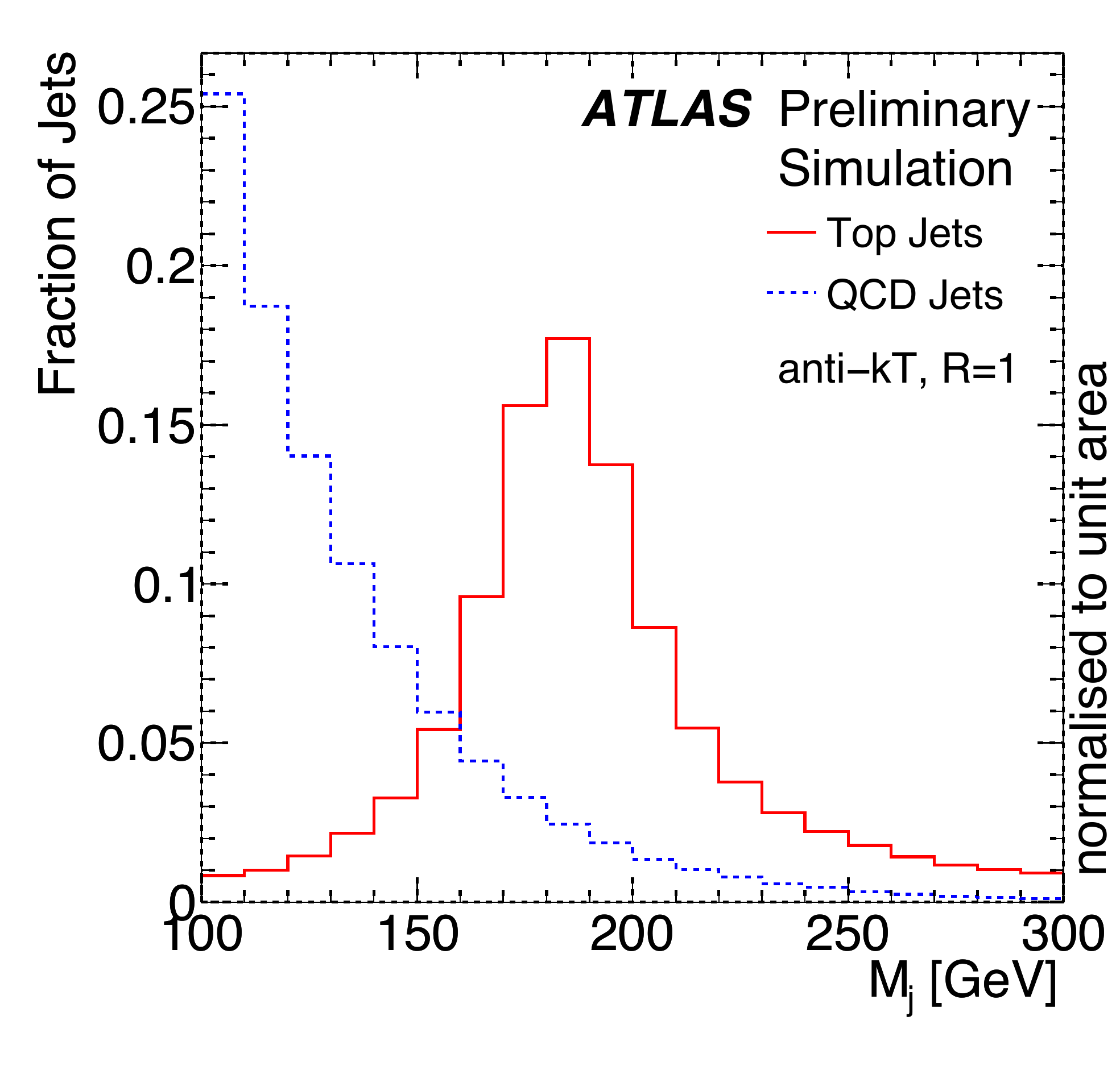} }
\subfigure[] { \includegraphics[width=0.23\textwidth]{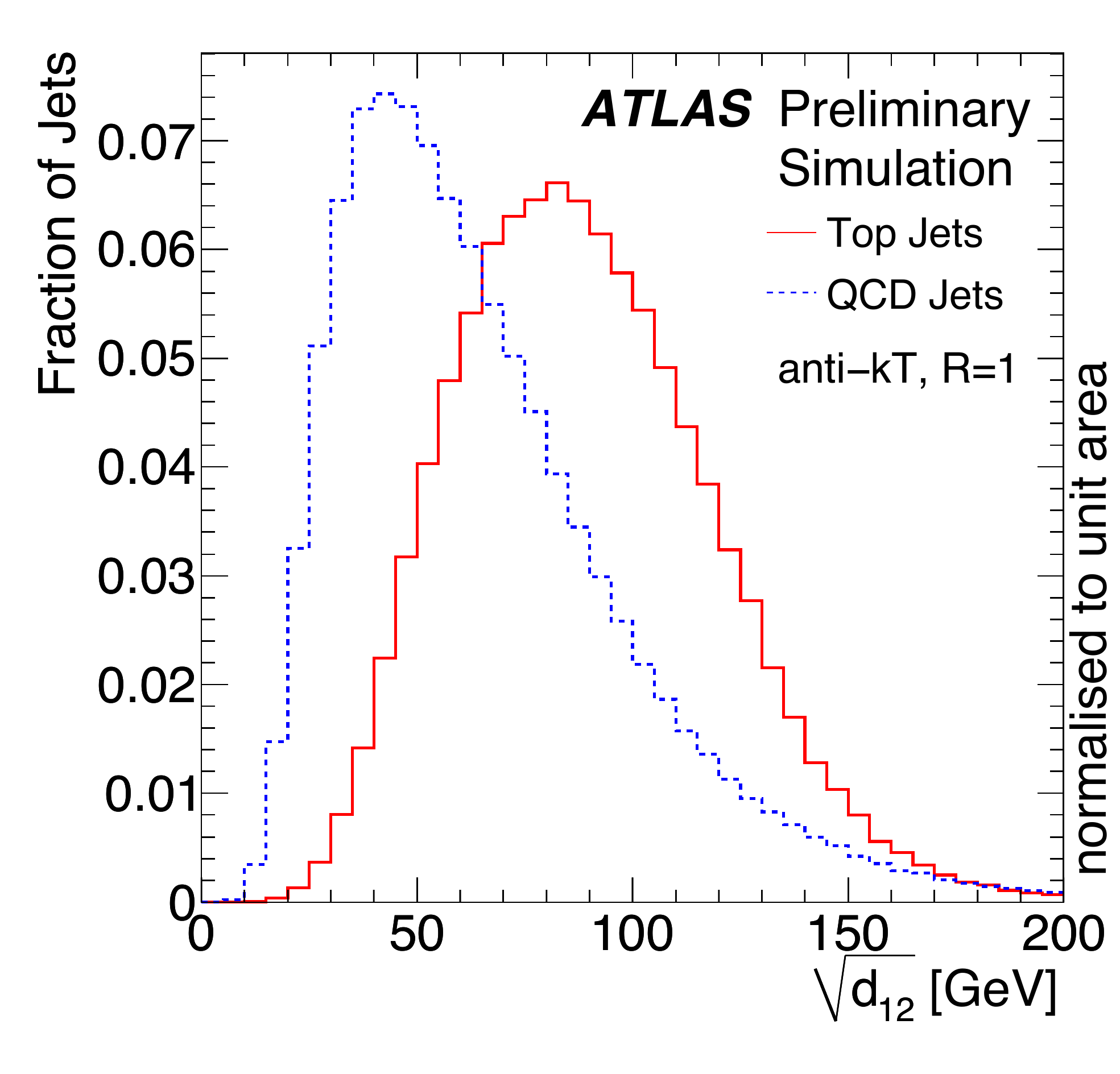} }
\subfigure[] { \includegraphics[width=0.23\textwidth]{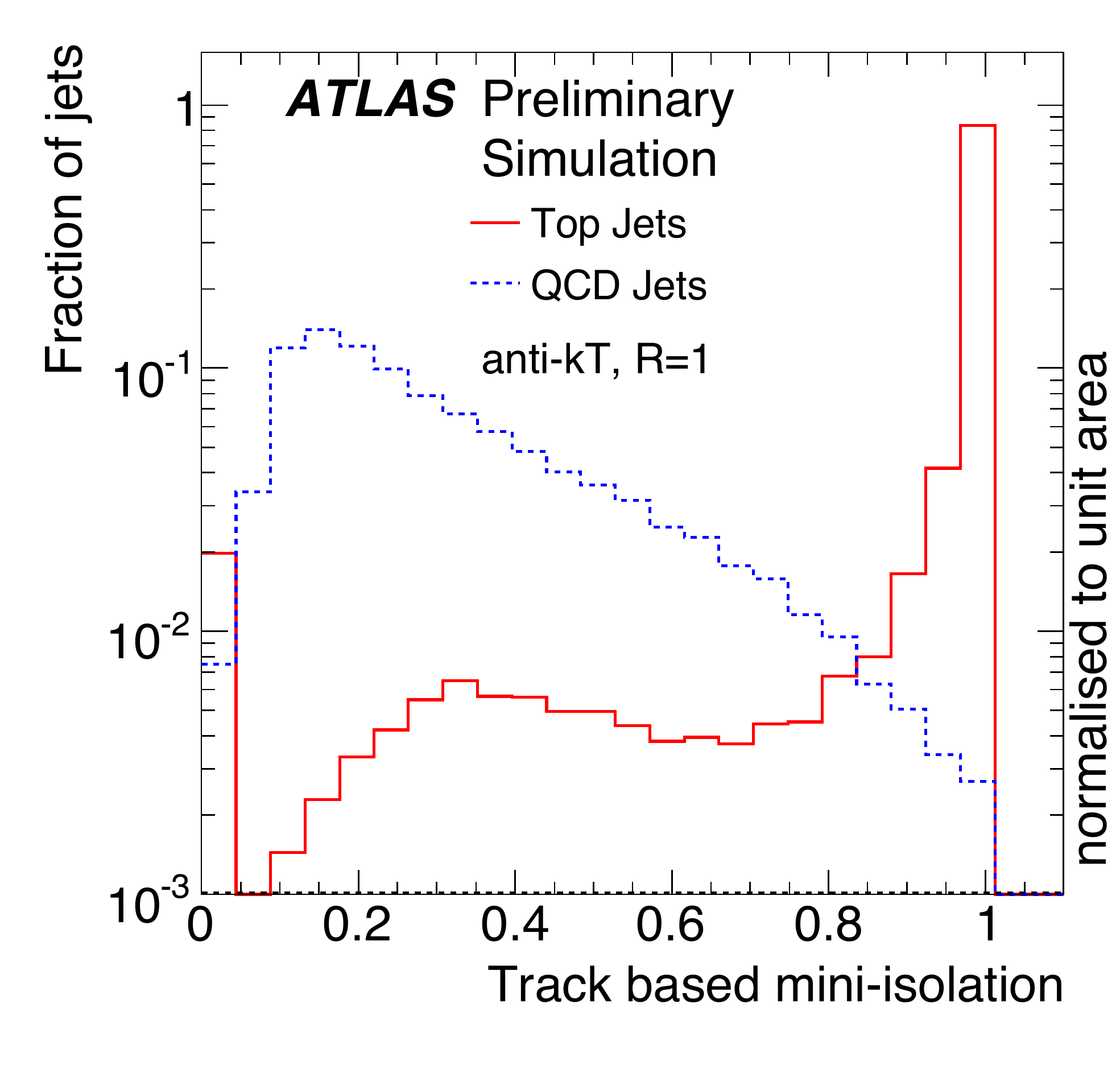} }
\subfigure[] { \includegraphics[width=0.23\textwidth]{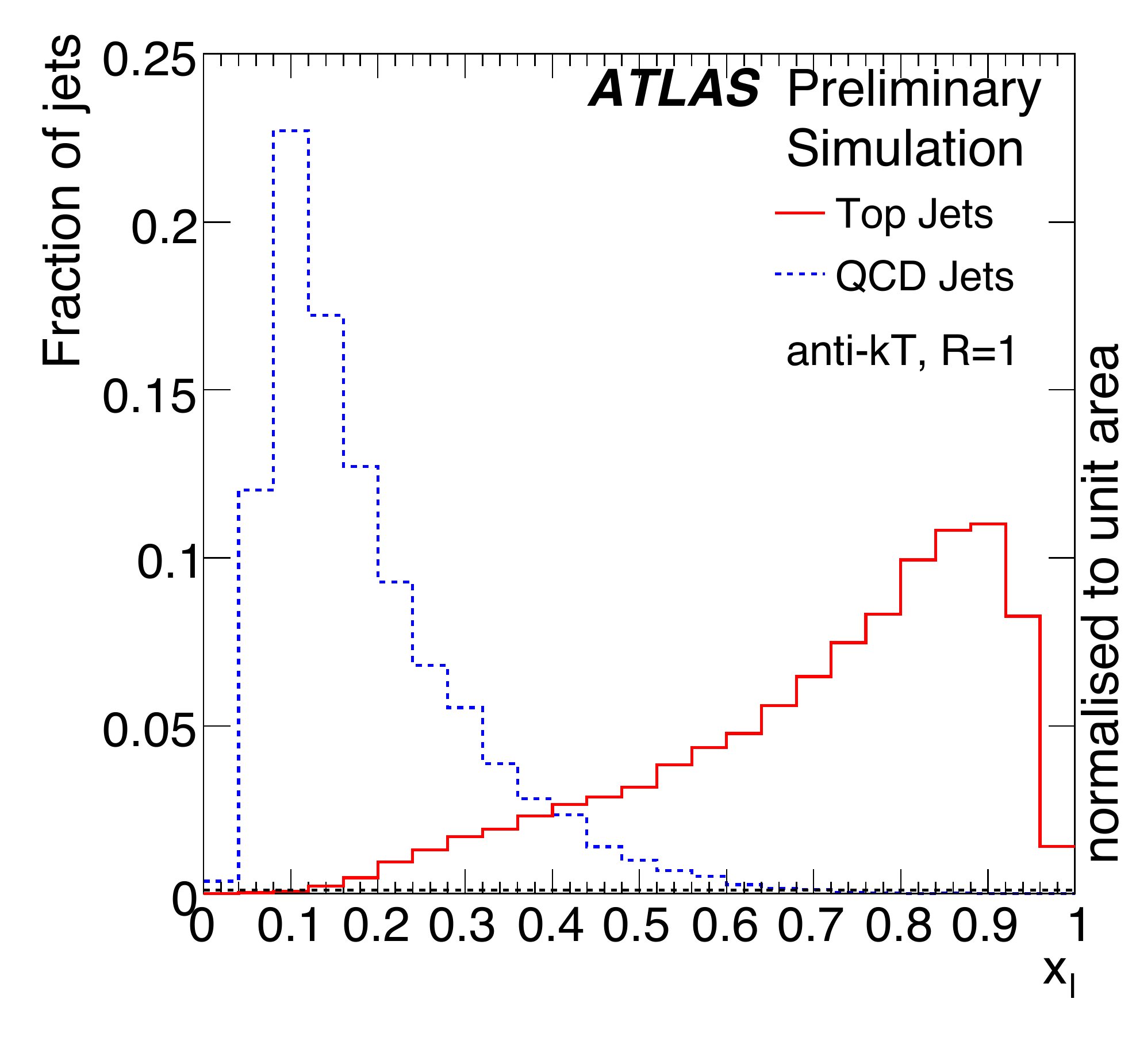} }
\caption{Observables probing jet substructure. a) Jet mass. b) First $k_\perp$ splitting scale. c) Lepton isolation in a cone whose size is a function of the lepton $p_T$. d) Fraction of the invariant visible mass carried away by the lepton.  }
\label{substructure}
\end{figure}


The high transverse momentum of the leptonically decaying top quark causes the lepton from the $W$ decay to be embedded in the jet. As a result, the traditional lepton isolation requirements become inefficient. One therefore needs to disentangle this signal from soft leptons in jets originating from the decay of $B$- and $D$-hadrons in heavy flavor QCD jets. This can be achieved by using discriminant observables to probe the presence of a hard lepton in the jet. Figure~\ref{substructure}.c,d shows distribution of such observables for boosted semi-leptonic top candidates.\\


For hadronically decaying top quarks in the mono-jet topology, the decay products are fully merged and reconstructed as a single fat jet. The challenge here is to disentangle top mono-jets from QCD high-$p_T$ jets. The jet mass (Figure~\ref{substructure}.a) is a natural observable to do so but it does not probe the jet hard substructure that we expect from a 3 body decay. To achieve this, the hierarchical nature of the $k_\perp$ jet algorithm is exploited by reclustering the initial jet's constituents with the $k_\perp$ algorithm. The last and penultimate stages of this process correspond on average to the merging of the top quark decay products and hence jet substructure can be probed via the first few $k_\perp$ splitting scales (Figure~\ref{substructure}.b).   

\section{CONCLUSION}
Three complementary algorithms for the reconstruction of the $t \bar t$ invariant mass spectrum
have been developed and their performance evaluated on fully simulated events. Two
adaptations of classical top reconstruction algorithms allow for high signal efficiency even in
the TeV regime ( $\sim 18$\% and 5\% in the $m=1-2$ TeV range for the minimal and full
reconstruction approaches respectively) . The mono-jet approach has been shown to be
efficient down to $m_{t\bar t} = 1$ TeV, with a signal efficiency of $\sim 9$\% (15\%) at $m=1$ TeV (2 TeV). Thus, the very challenging mass region where different topologies coexist is covered. If no deviation from the Standard Model is observed, a 95\% C.L. limit of $\sigma \times$ BR($X \rightarrow t \bar t$) = 3
pb is expected for a resonance mass of 1 TeV after 200 pb$^{-1}$ at center-of-mass energy of 10
TeV. Approximately the same sensitivity for $m=1$ TeV is expected for 1 fb$^{-1}$ of data at 7 TeV.

\end{document}